\documentclass[review]{elsarticle}

\makeatletter
\def\ps@pprintTitle{%
 \let\@oddhead\@empty
 \let\@evenhead\@empty
 \def\@oddfoot{\centerline{\thepage}}%
 \let\@evenfoot\@oddfoot}
\makeatother

\usepackage{epstopdf}
\usepackage{multicol}
\usepackage[english]{babel}
\usepackage{lipsum}
\usepackage{amsmath,array,graphicx}
\usepackage{amssymb}
\usepackage{sidecap}
\sidecaptionvpos{figure}{t}
\usepackage[verbose]{placeins}
\usepackage{subfigure}
\usepackage{placeins}
\usepackage{multirow}
\usepackage{lineno,hyperref}
\usepackage{textcomp}
\usepackage[usenames, dvipsnames]{color}

\journal{Journal of Powder Technology}

\begin{document}

\begin{frontmatter}

\title{Stretching the limits of dynamic and quasi-static flow testing on limestone powders}

\author[MSMaddress]{Hao Shi\corref{mycorrespondingauthor}}
\cortext[mycorrespondingauthor]{Corresponding Author}
\ead{h.shi-1@utwente.nl}

\author[Geoffroyaddress]{Geoffroy Lumay}
\ead{geoffroy.lumay@uliege.be}

\author[MSMaddress]{Stefan Luding}
\ead{s.luding@utwente.nl}

\address[MSMaddress]{Multi Scale Mechanics, TFE, ET, MESA+, University of Twente, 7500 AE Enschede, The Netherlands}

\address[Geoffroyaddress]{GRASP Laboratory, CESAM Research Unit, University of Li\`ege, Belgium}

\begin{abstract}

Powders are a special class of granular matter due to the important role of cohesive forces. The flow behavior of powders depends on the flow states and stress and is therefore difficult to measure/quantify with only one experiment. In this study, the most commonly used characterization tests that cover a wide range of states are compared: (static, free surface) angle of repose, the (quasi-static, confined) ring shear steady state angle of internal friction, and the (dynamic, free surface) rotating drum flow angle are considered for free flowing, moderately and strongly cohesive limestone powders.
 
The free flowing powder gives good agreement among all different situations (devices), while the moderately and strongly cohesive powders behave more interestingly. Starting from the flow angle in the rotating drum and going slower, one can extrapolate to the limit of zero rotation rate, but then observes that the angle of repose measured from the heap is considerably larger, possibly due to its special history. When we stretch the ring shear test to its lowest confining stress limit, the steady state angle of internal friction of the cohesive powder coincides with the flow angle (at free surface) in the zero rotation rate limit.

\end{abstract}


\end{frontmatter}


\section{Introduction}
\label{intro}

\par
Granular media are a collection of discrete solid particles interacting through dissipative contact forces; their natural discontinuity poses many challenges for both academia and industry in understanding their bulk behavior \cite{jaeger1996granular}. One of the challenges when dealing with granular media in processes is the characterization of these materials. While the characterization at the scale of the grains (size and shape distribution, ...) is sometimes difficult, the macroscopic characterization (flow, packing fraction, tendency to segregate, ...) is also tricky and a wide variety of tests are available \cite{schwedes2003review}. 
\par
Since decades, granular media have been subject to many fundamental studies, ranging from static to flowing conditions, from hard to soft particles, and from low to very high stresses. Micro-mechanical studies of granular materials give an essential understanding of their macro-scale behavior. For example, at micro or meso scale, the study by Radjai et al. \cite{radjai1998bimodal} classifies the contacts into subnetworks of strong and weak contacts: the anisotropic shear stress of granular materials is primarily carried by the strong contacts. This method offers insight into the micro structure change from the contact origin but has its limitations for studying real life materials, e.g., limestone powders, especially the very fine ones which are strongly cohesive. The cohesion at micro scale can not be easily scaled up due to the complexity at meso scale \cite{tomas2007adhesion,berger2016scaling,shi2018effect}, and there are still little focuses on the interesting behaviour of cohesive granular flow.

\par
At macroscopic scale, from the perspective of granular flow, researchers have investigated different dynamic flow configurations like plane shear cells, Couette cells, silos, flows down inclined planes, or avalanches on piles and in rotating drums \cite{midi2004dense,jop2006constitutive,pouliquen2006flow,lumay2006linking,forterre2008flows,schall2010shear,lumay2010flow,lumay2012measuring,jarray2019wet}, where the granular materials are usually under very low or even free surface conditions. From the perspective of material characterization, researchers have developed various element tests in the lab to quantify the bulk responses of granular materials under specific stress/strain conditions. Element tests are (ideally homogeneous) macroscopic laboratory tests in which the force (stress) and/or displacement (strain) path are controlled. One of the most widely performed element tests in both industry and academia is the shear test in various designs \cite{casagrande1936determination, jenike1964storage,schwedes1979vergleichende,schulze1994entwicklung,shibuya1997interpretation,schwedes2003review,schulze2008powders,zafar2015comparison,shi2018effect}, where a granular sample is sheared until failure is reached and the material starts to flow. The shear zone is pre-defined by the device design, and the shear failure is forced in a specific physical location. Another common element tests are the uni-axial compression tester \cite{russell2014influences, thakur2014experimental, imole2016slow} where the lateral stress ($\lambda$-test) is more challenging as a bi-axial shear box \cite{morgeneyer2003can,morgeneyer2003investigation,feise1995investigation}. All these element tests are done in static to quasi-static flow regimes, with the stress applied  usually above a few hundred pascals, while the granular flow tests mentioned above are normally carried out under more dynamic, and lower stress conditions. 

\par
In parallel to the classical shear cell test, different methods are commonly used to measure powder flow behavior: the angle of repose \cite{zhou2002experimental,ileleji2008angle}, the Hausner ratio \cite{grey1969hausner,traina2013flow}, flow in rotating drums \cite{liu2005experimental,pirard2009motion,nalluri2010flowability}, flow through orifices \cite{ahn2008experimental}, and powder rheometers with rotating blades inspired by liquid rheometers \cite{freeman2007measuring,madariaga2009characterization}. Different versions of each test exist from the simple manual \cite{european2010european} to automatic versions \cite{lumay2012measuring,traina2013flow}.    
\par
Some of the flow tests are dynamic while others are static or quasi-static. Moreover, some tests are conducted with a free powder surface, whereas others are performed under confinement. Finally, both flow and stress fields are depending on the geometry of the tester. The link between different tests is mostly missing and represents a great challenge. Therefore, in this study, we explore the connection between two types of tests by stretching their limits: explore the dynamic rotating drum towards very low rotation rate, hence going to the quasi-static regime; and bring the quasi-static ring shear tests towards very low confining stresses, thus approaching the stress conditions in the dynamic drum test.
\par
This study is structured as follows. Section \ref{sec:material} introduces the limestone materials; the description of the experimental devices and the test procedures are given in section \ref{sec:experimental}. 
Section \ref{sec:results} is devoted to the discussion of experimental results and bridging between dynamic and quasi-static tests and covering a wide stress range. Conclusions 
and outlook are presented in section \ref{sec:conclusionandoutlook}.

\section{Material description and characterization}
\label{sec:material}

Limestone powder is a widely used powder in fields ranging from construction to automotive industries. In this work, eight grades of pre-sieved limestone powder  under the commercial name Eskal (KSL Staubtechnik GmbH, Germany) are used. Eskal has been used as a reference powder for standard testing \cite{shi2018effect} and calibration of equipment in powder technology, for instance, shear testers \citep{feise1998review, zetzener2002relaxation}, and optical sizing systems due to its favourable physical properties: high roundness, low porosity and an almost negligible sensitivity to humidity and temperature changes, which allows to avoid sample pretreatment.

Each grade of the Eskal series is milled and then sieved to ensure a certain particle size distribution. Three grades of Eskal are chosen specifically from the experience in a previous study \cite{shi2018effect}: fine/cohesive Eskal300 ($d_{50}$ = 2.22 \textmu m), slightly cohesive Eskal15 ($d_{50}$ = 19 \textmu m) and coarse/free-flowing Eskal150 ($d_{50}$ = 138 \textmu m). The details of their physical properties are summarized in Table \ref{tablematerialEskal}.

\begin{figure}[!ht]
 \centering
\includegraphics[scale=0.45,angle=0]{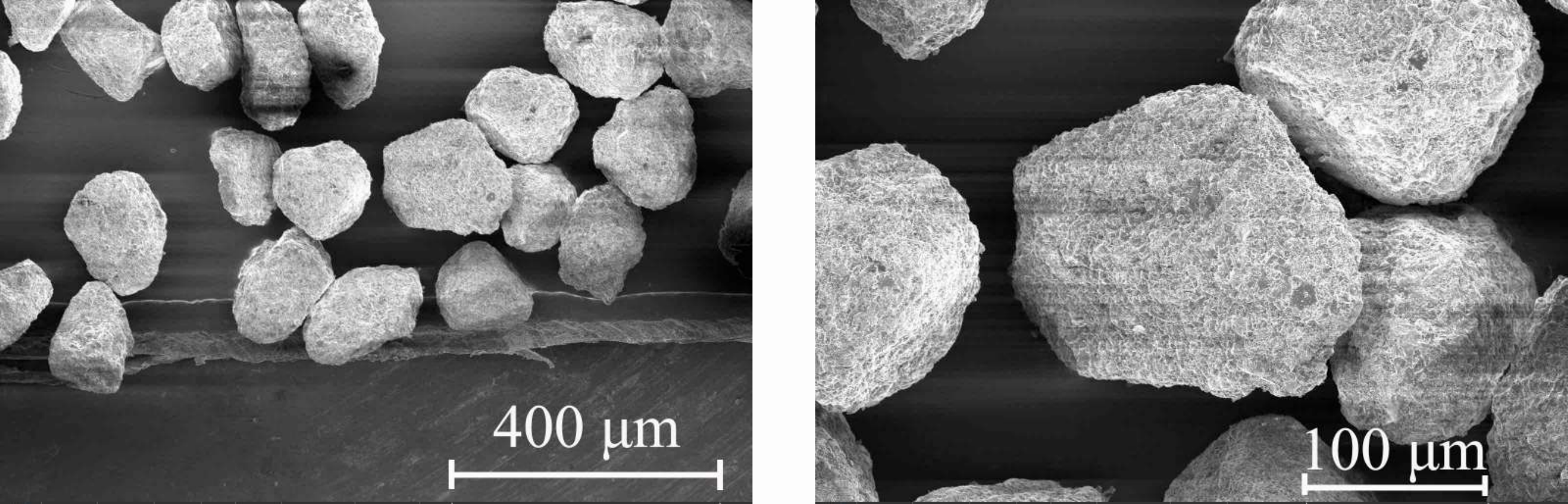}\label{eskal150-sem}
\caption{SEM images of Eskal150 ($d_{50}$ = 138 \textmu m) in two different magnifications: 185x (left) and 502x (right).}
\label{fig-sec2:eskal150-sem}
\end{figure}

\begin{figure}[!ht]
 \centering
\includegraphics[scale=0.45,angle=0]{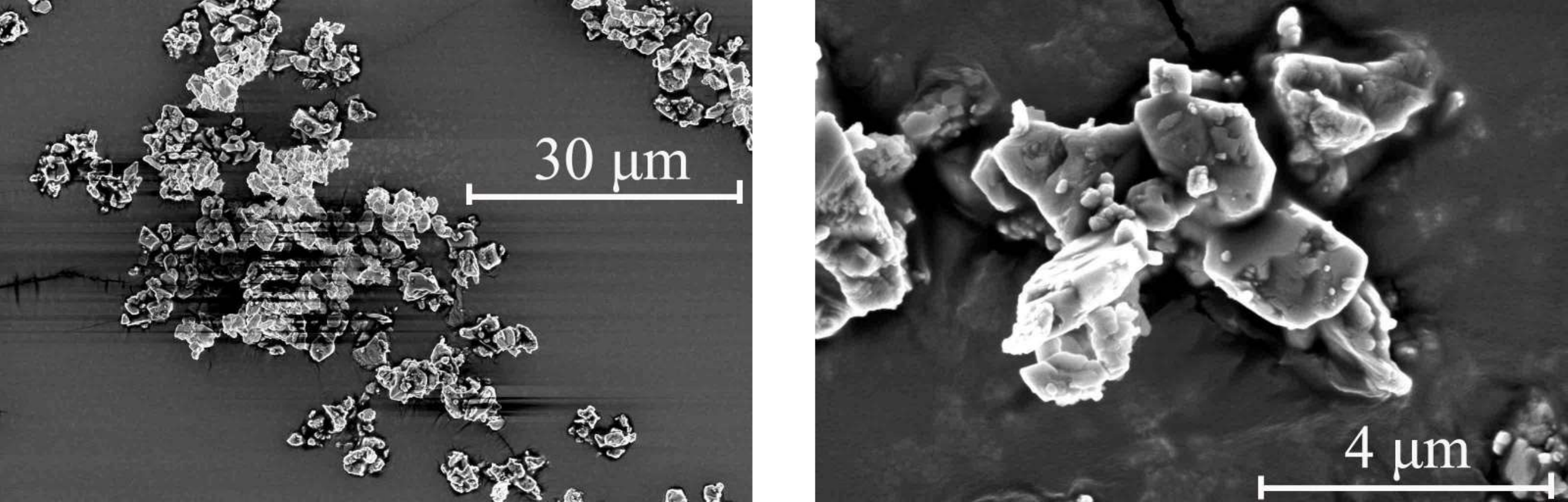}\label{eskal300-sem}
\caption{SEM images of Eskal300 ($d_{50}$ = 2.2 \textmu m). Magnifications: 2500x (left) and 15000x (right).}
\label{fig-sec2:eskal300-sem}
\end{figure}

The aspect ratio, shape and morphology of Eskal 150 and Eskal 300 are analyzed by means of Scanning Electron Microscope (SEM) imaging. Materials were sputtered with silver and investigated with a field emission instrument (Helios G4 CX, FEI Deutschland GmbH, Germany) with an EDX detector, applying an acceleration voltage of 5 kV and a working distance of 4 or 6 mm. Different magnifications between 185x and 15000x were applied. Figures\ \ref{fig-sec2:eskal150-sem} and \ref{fig-sec2:eskal300-sem} show the SEM images of Eskal150 and Eskal300, respectively. 
In Fig.\ \ref{fig-sec2:eskal150-sem}, we see that all the Eskal150 primary particles have similar shapes (left) and rough surfaces (right), and every particle is clearly distinguished/separated from the others. In contrast, for Eskal300 in Fig.\ \ref{fig-sec2:eskal300-sem} (left), we observe some clusters of primary particles, and the size of clusters is typically around 10 to 20 \textmu m, which is about 5 to 10 times the median particle size of Eskal300. When we zoom into a smaller scale, focusing on one single cluster as shown in Fig.\ \ref{fig-sec2:eskal300-sem} (right), we see even smaller fines ($<$ 1 \textmu m) sticking on the surface of primary particles. Moreover, the shapes of Eskal300 particles are more irregular than Eskal150 particles.

\section{Experimental Setup}
\label{sec:experimental}

\begin{figure}[!ht]
\centering
\includegraphics[scale=0.26,angle=0]{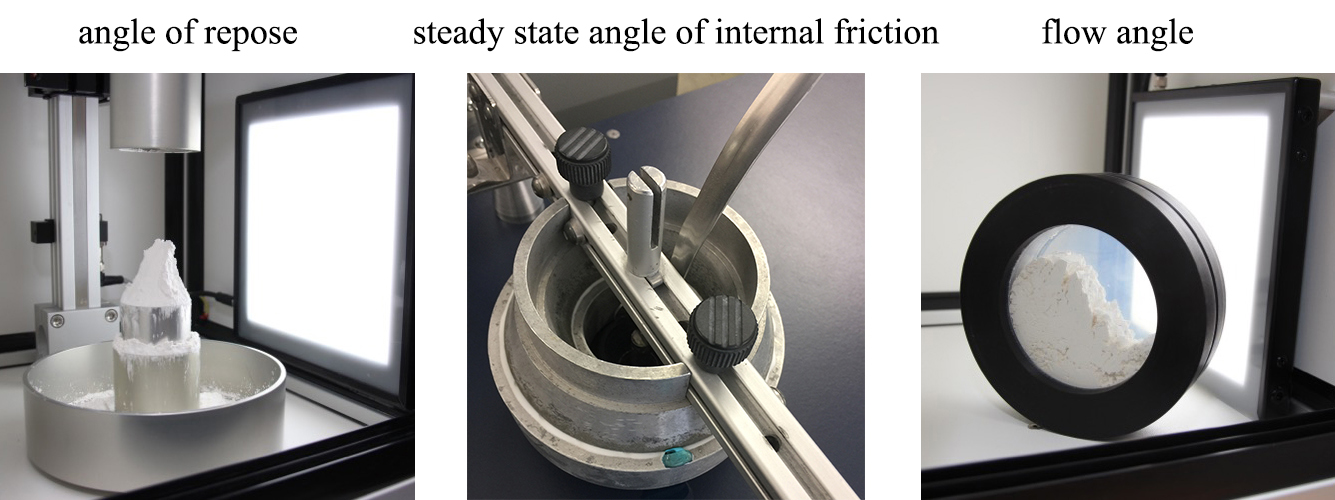}
\caption{Left: GranuHeap for measuring angle of repose; middle: the Schulze ring shear tester RST-01 for measuring steady state angle of internal friction; right: GranuDrum for measuring the flow angle.}
\label{fig-sec3:joint-device}
\end{figure}

In this study, we combine three experimental devices: GranuHeap (angle of repose), Schulze ring shear tester (steady state angle of internal friction), and GranuDrum (flow angle), to perform measurements in both static and dynamic regimes. The details of each setup are shown in Fig.\ \ref{fig-sec3:joint-device} and will be explained in the following.

\subsection{GranuHeap - Static Free Surface}
\label{sub-sec:granule-heap-intro}

The angle of repose test has been widely used since 1943 in the particle and powder community. Al-Hashemi and Al-Amoudi presented a very wide review on different methods to obtain the angle of repose both experimentally and numerically \cite{al2018review}. The GranuHeap instrument \cite{lumay2012measuring} is an automated repose angle measurement device based on image processing and uses the principle of hollow cylinder method categorized in \cite{al2018review}. A powder heap is created on a cylindrical support to be analyzed by image processing. The geometry of the measurement cell and a typical heap are presented in Fig.\ \ref{fig-sec3:joint-device} (left). In order to obtain reproducible results, an initialization tube with an internal diameter equal to the circular support is installed on the support. After filling the initialization tube by hand with a fixed volume of powder (100 ml in the case of the present study), the tube moves up at a constant speed of 5 mm/s. Thereby, the powder is flowing from the tube to form a heap on the cylindrical support, which is then evaluated by image analysis. A controlled rotation of the support allows obtaining different heap projections. In the present study, 16 images separated by a rotation angle of 11.25$^\circ$ were recorded. A custom image recognition algorithm determines the position of the powder/air interface. The angle of repose $\phi_{sta}$ refers to the angle of the isosceles triangle with the same projected surface as the powder heap. The isosceles triangle corresponds to the ideal heap shape. The lower the repose angle is, the better the powder flowability is \cite{lumay2012measuring}. A static cohesive index $\sigma_{sta}$ can be also measured from the interface irregularities (not shown in the present study).

\subsection{Schulze Ring Shear Tester - RST-01 - Quasi-static Confined Surface}
\label{sub-sec:schulze-rst01-intro}

Shear testers are used for powder characterization since decades. The Schulze rotational ring shear tester (1994) is one of the most widely used testers and it is semi-automated.
The Schulze ring shear tester (RST-01) operates connected to a personal computer running a control software that allows the user to obtain, among other things, yield loci and wall yield loci. The ring-shaped (annular) bottom ring of the shear cell contains the bulk solid specimen. An annular-shaped lid is placed on top of the bulk solid specimen and it is fixed at a cross-beam (Fig.\ \ref{fig-sec3:joint-device}, middle).
A normal force, $F_N$, is exerted on the cross-beam in the rotational axis of the shear cell and transmitted through the lid onto the specimen, i.e, a controlled normal stress is applied to the bulk solid. In order to allow small confining stress, the counterbalance force, $F_A$, acts in the centre of the cross-beam, created by counterweights and directed upwards, counteracting the gravity forces of the lid, the hanger and the cross-beam. Shearing of the sample is achieved by rotating the bottom ring with an angular velocity $\omega$, whereas the lid and the cross-beam are prevented from rotation by two tie-rods connected to the cross-beam. Each of the tie-rods is fixed at a load beam, so that the forces, $F_1$ and $F_2$, acting on the tie-rods can be measured. The bottom of the shear cell and the lower side of the lid are rough in order to prevent sliding of the bulk solid on these two surfaces. Therefore, rotation of the bottom ring relative to the lid creates a shear deformation within the bulk solid. Through this shearing the bulk solid is deformed, and thus a shear stress $\tau$ develops, proportional to the forces on the tie-rods ($F_1$ + $F_2$). All the tests performed here follow the procedure as in the ASTM standard \cite{astm2008standard}.

Typical confining stresses used in the shear cell tests are between 1 and 10 kPa. However, this is too high compared to the pressure range of free or nearly free surface. Thus, in order to explore the low confining stress regime, we employ the pre-shear normal stresses down to the device's lowest limit: 2, 1.5, 1.0, 0.8, 0.6, 0.4, 0.2 and 0.1 kPa. For cohesive Eskal300, we could apply the pre-shear normal stresses down to 0.1 kPa, whereas for free-flowing Eskal150, the minimum is at 0.2 kPa. And in order to achieve very low pre-shear normal stress in RST-01, we use a special shear cell lid made of PVC instead of aluminium, which has a lower self weight of the lid and allows to apply very low stress. However, at the lowest stresses the pre-consolidation becomes questionable and the output is not representative. For each pre-shear normal stress, we performed three runs, with every time a fresh sample, in order to investigate repeatability. In all tests presented here, the shear velocity is kept constant (1 mm/min) as default to ensure that the shearing is within the quasi-static regime.
A typical testing procedure is as follows: first vertically compress the sample to the predefined pre-shear normal stress value, e.g. 1 kPa and deploy shear, the control software will wait until the shear stress is almost constant then stop the shearing. Then the first pre-shear point is obtained. The normal stress is kept at pre-shear and the lid will rotate backwards to reach a zero shear stress state, then the normal stress will reduce to the first shear point, e.g. 0.4 kPa and continue shearing. After a peak failure in the shear stress is detected, the first shear cycle is finished and thus the first shear point is obtained. The software/program will continue this pre-shear then shear procedure until all the shear points are measured. A more detailed explanation of different procedures are given in \cite{shi2018effect,schulze2008powders} and will not be further addressed here for the sake of brevity.

\subsection{GranuDrum - Dynamic Free Surface}
\label{sub-sec:rotating-drum-intro}

The GranuDrum instrument \cite{lumay2012measuring} is an automated powder flowability measurement technique based on the rotating drum geometry, which characterizes materials in the dynamic flowing regime with a free surface. A horizontal cylinder with vertical glass side walls (called drum) is half filled with the sample of powder. For the present study, the drum rotates around its horizontal axis of symmetry at rotating speeds from 2 RPM to 10 RPM (increase sequence) and we do not analyze the flow during the rotating speed decrease sequence. A CCD camera takes snapshots (50 images separated by 0.5s) at each angular velocity (see Fig.\ \ref{fig-sec3:joint-device} right). The air/powder interface is detected on each snapshot with an edge detection algorithm. Afterward, the average interface position and the fluctuations around this average position are computed. Then, for each rotating speed, the dynamic friction (flow) angle $\phi_{dyn}$ is measured at the center of the average interface position. A dynamic cohesive index $\sigma_{dyn}$ can be also measured from the interface fluctuations (not shown in the present study).

In order to compare the confined surface ring shear test to the free surface GranuHeap and GranuDrum, we proposed a simple method to estimate the (effective) confining stress on flowing powders in both GranuHeap and GranuDrum tests by two principles: single particle layer $h_0$ and effective flowing depth of the rotating drum $h$. The first one represents the effective pressure induced by a single layer of primary particles, which can be correlated to the static GranuHeap test. In the static situation, one expects the flow depth close to the free surface to be the same order of magnitude of the particle diameter.

The effective flowing depth is valid only for the case of the rotating drum and given by the ratio between the actual flowing depth $h$ and the radius $r$ of the drum $h_{eff} = h/r$. The flowing depth of non-cohesive granular materials in a rotating drum depends on the rotating speed and on the ratio between the drum diameter and the grain diameter \cite{felix2007granular}. For cohesive powders, the flowing depth increases with the cohesiveness \cite{brewster2009effects}, the powder particles will form agglomerates/aggregates during the flow/movement, but those agglomerates/aggregates are not fully stable, they might break and reform again. It is almost impossible to get an accurate measurement of the depth of the flowing layer for our cohesive Eskal300. Therefore, instead of giving an estimation of the flowing depth of our cohesive Eskal300, we use a depth range: 1\% to 20\% of the drum radius, which covers almost all the possible depths of cohesive powder flows in a rotating drum \cite{brewster2009effects}. Then, the effective confining stresses are evaluated at different depths $h$ inside the powder bed considering the hydro-static pressure $\sigma = \rho_{bulk} g h$, where $\rho_{bulk}$ is the powder bulk density and $g$ is gravitational acceleration.

\FloatBarrier

\section{Results and Discussion}
\label{sec:results}

\subsection{Static Granular Heap}
\label{subsec:results-heap}

\begin{figure}[!ht]
 \centering
\includegraphics[scale=0.32,angle=0]{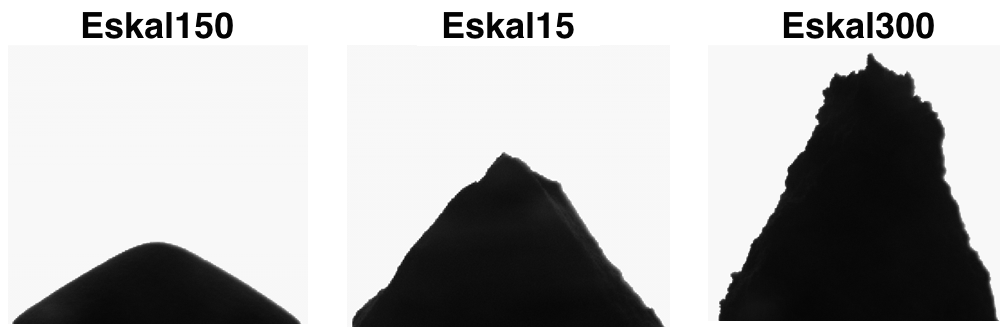}\label{picturesHeap}
\caption{Typical heaps obtained with Eskal150, Eskal15 and Eskal300.}
\label{fig-sec4:picturesHeap}
\end{figure}

Fig.\ \ref{fig-sec4:picturesHeap} shows typical heaps obtained with Eskal150 (left) Eskal15 (middle) and Eskal300 (right).
The cohesive Eskal300 powder forms a strongly irregular heap with a high static friction (repose) angle ($\phi_{sta} = 69.1 \pm 1.9 ^{\circ}$) due to the influence of cohesion between particles. At the opposite, the heap obtained with Eskal150 has an almost conical shape with a low angle of repose ($\phi_{sta} = 33.0 \pm 0.1 ^{\circ}$). While the heap obtained with Eskal15 has a very similar conical shape as Eskal150 with small irregular shape and the obtained angle of repose of Eskal15 stays between the angles of repose of Eskal150 and 300 ($\phi_{sta} = 52.6 \pm 1.4 ^{\circ}$).
It has been widely investigated that even for one method, there are still difficulties in the repeatability and reproducibility, due to human/operator influences inside a single lab or at different labs \cite{al2018review}. In the current study, the repose angle measurement of each Eskal powder has been repeated four times with fresh samples to obtain a representative mean value with the rather good repeatability (2.7\% for cohesive Eskal300, 2.6\% for moderately cohesive Eskal15 and 0.3\% for free flowing Eskal150). Moreover, in each single measurement, the stably formed heap was rotated slowly to take 16 pictures at different viewing angles from the side of the heap and then averaged to obtain the final value.

\FloatBarrier
\subsection{Quasi-Static Ring Shear Tester}
\label{subsec:results-ring}

\begin{figure}[!ht]
 \centering
\includegraphics[scale=0.4,angle=0]{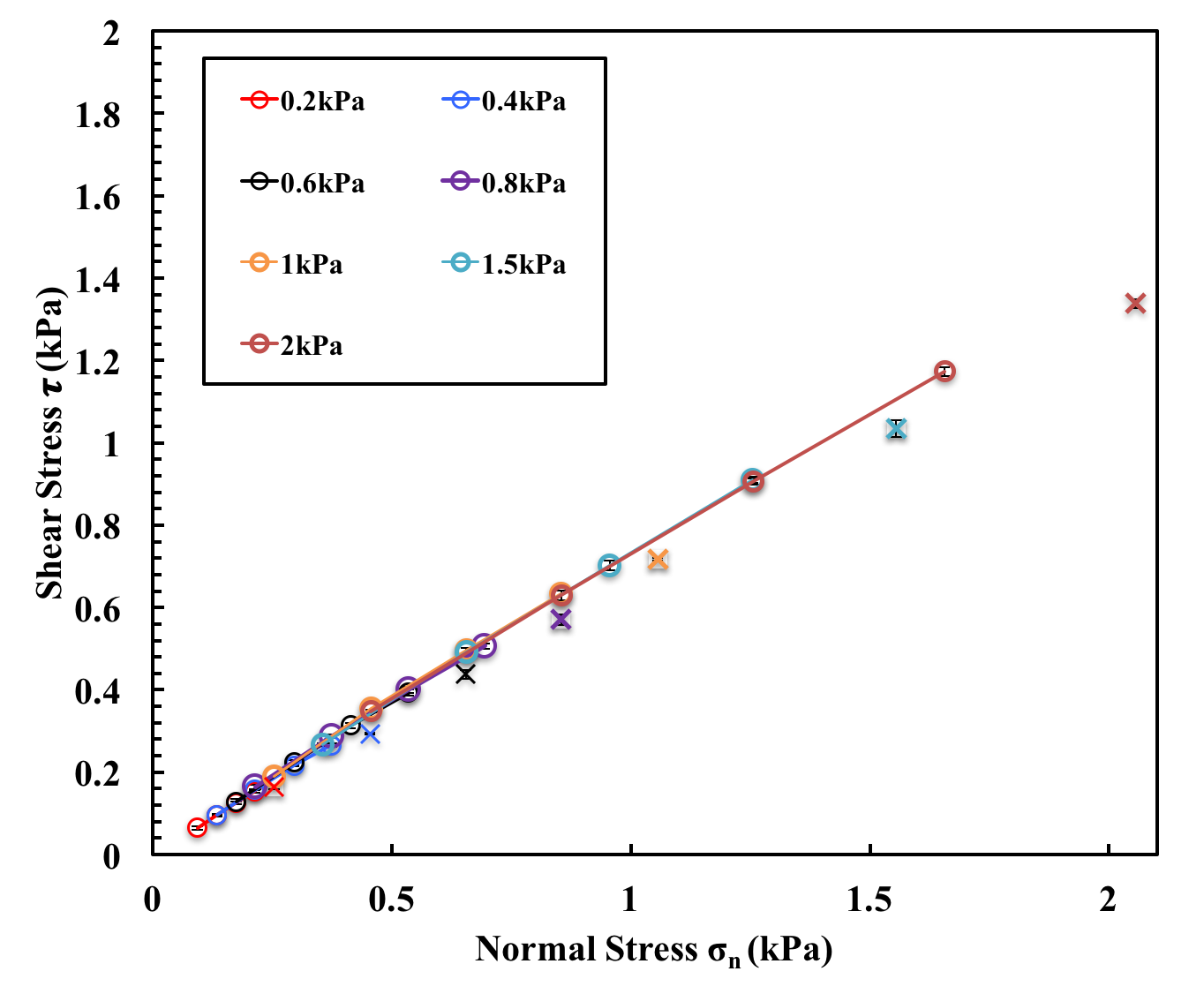}\label{eskal150-low-YL}
\caption{Yield locus (shear stress versus normal stress) of Eskal150 ($d_{50}$ = 138 \textmu m) using RST-01.pc. The pre-shear normal stress is kept between 0.2 and 2 kPa. Different colours indicate different pre-shear normal stresses. Points with and without lines are shear and pre-shear points, respectively. Lines are only guides to the eye.}
\label{fig-sec5:eskal150-low-YL}
\end{figure}

In the low confining (normal) stress regime, we first look at the yield loci at different pre-shear stresses (0.2 to 2 kPa). Each yield locus is measured with 3 fresh samples to acquire the standard deviations. The yield loci for Eskal150 ($d_{50}$ = 138 \textmu m) are shown in Fig.\ \ref{fig-sec5:eskal150-low-YL} with different pre-shear stresses indicated by different colours. With increase in pre-shear normal stress, all the yield loci  collapse on a single curve. This is expected for free flowing powder, where the flow behaviour is not sensitive to the pre-shear confining stress. The pre-shear points stay consistently lower than the corresponding yield loci. However, the difference between the pre-shear points and yield loci increases with the pre-shear normal stress. Both pre-shear and shear points show very good repeatability with maximum standard deviations around symbol size. We only manage to measure representative yield locus of Eskal150 down to 0.2 kPa pre-shear normal stress, while the data measured at lower stress levels are not reliable. Note that we have also measured the yield loci at 3, 4 and 5 kPa (see Ref. \cite{garcia2019tensile}), but for the sake of brevity, the data are not shown here, since they all follow the trend of low stress levels data.

\begin{figure}[!ht]
 \centering
\includegraphics[scale=0.4,angle=0]{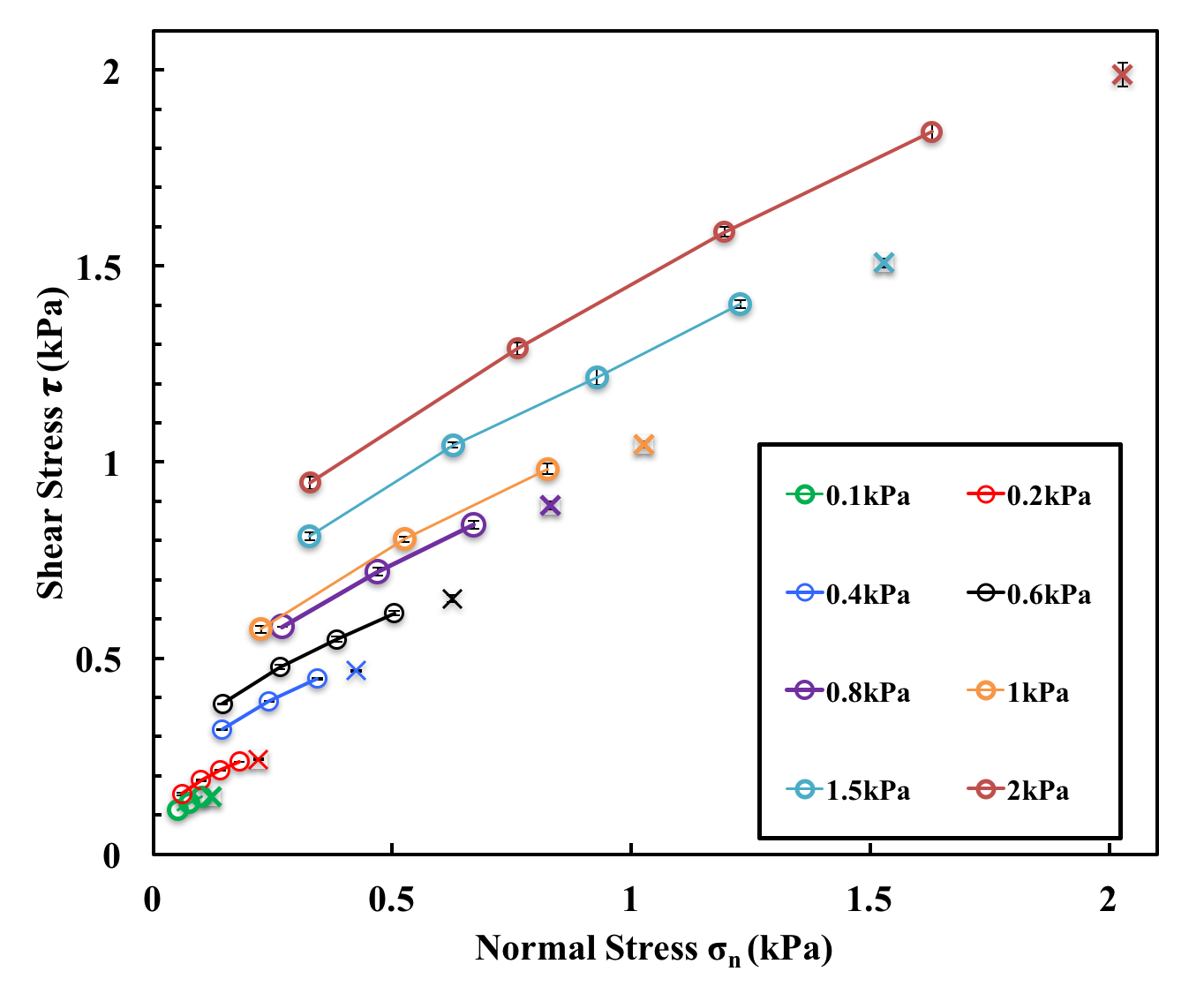}\label{eskal300-low-YL}
\caption{Yield locus (shear stress versus normal stress) of Eskal300 ($d_{50}$ = 2.2 \textmu m) using RST-01.pc. The pre-shear normal stress is kept between 0.1 and 2 kPa. Different colours indicate different pre-shear normal stresses. Points with and without lines are shear and pre-shear points, respectively. Lines are only guides to the eye.}
\label{fig-sec5:eskal300-low-YL}
\end{figure}

For the cohesive Eskal300 ($d_{50}$ = 2.2 \textmu m), we measured the yield loci in the normal stress range between 0.1 and 2 kPa, and the data are shown in Fig.\ \ref{fig-sec5:eskal300-low-YL}. Unlike the free flowing Eskal150, the yield loci of Eskal300 move upwards with the increase of the pre-shear normal stress, which indicates the cohesive Eskal300 is sensitive to the pre-shear normal stress. The yield loci of Eskal300 show a convex curvature as clearly visible from the guide lines as studied in \cite{garcia2019tensile}. Similar to the case of Eskal150, the yield loci of Eskal300 show very good repeatability with maximum standard deviations around symbol size.
Furthermore, we have also included the steady state angle of internal friction of moderate cohesive Eskal15 at three pre-shear stress levels: 5, 20 and 35 kPa (data only shown in Fig.\ \ref{fig-sec5:drum-RST-fitting}) from our previous study \cite{shi2018effect} for the sake of validation.

\FloatBarrier
\subsection{Unifying the Static and the Dynamic States}
\label{subsec:results-drum}

\begin{figure}[!ht]
 \centering
\includegraphics[scale=0.35,angle=0]{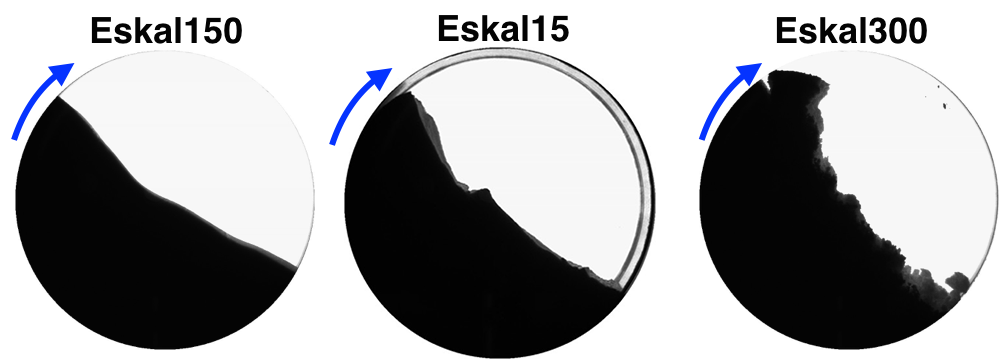}\label{picturesDrum}
\caption{Snapshots of typical flow inside the rotating drum with Eskal150, Eskal15 and Eskal300.}
\label{fig-sec4:picturesDrum}
\end{figure}

Following the same principle as in Figure \ref{fig-sec4:picturesHeap} with heaps, Figure \ref{fig-sec4:picturesDrum} shows the typical flowing patterns obtained in the rotating drum with Eskal150 (left), Eskal15 (middle) and Eskal300 (right). The free flowing Eskal150 shows a very smooth free surface with a slightly concave shape, while the cohesive Eskal300 gives a much rougher free surface with some clumps due to cohesion. For the sake of completeness, we also added here another slightly cohesive powder Eskal15 ($d_{50} = 19 $ \textmu m) which lays between Eskal150 and Eskal300. The surface of Eskal15 powder has the same concave shape as cohesive Eskal300, but much smoother (less clumps), which is expected.

\begin{figure}[!ht]
 \centering
\includegraphics[scale=0.36,angle=0]{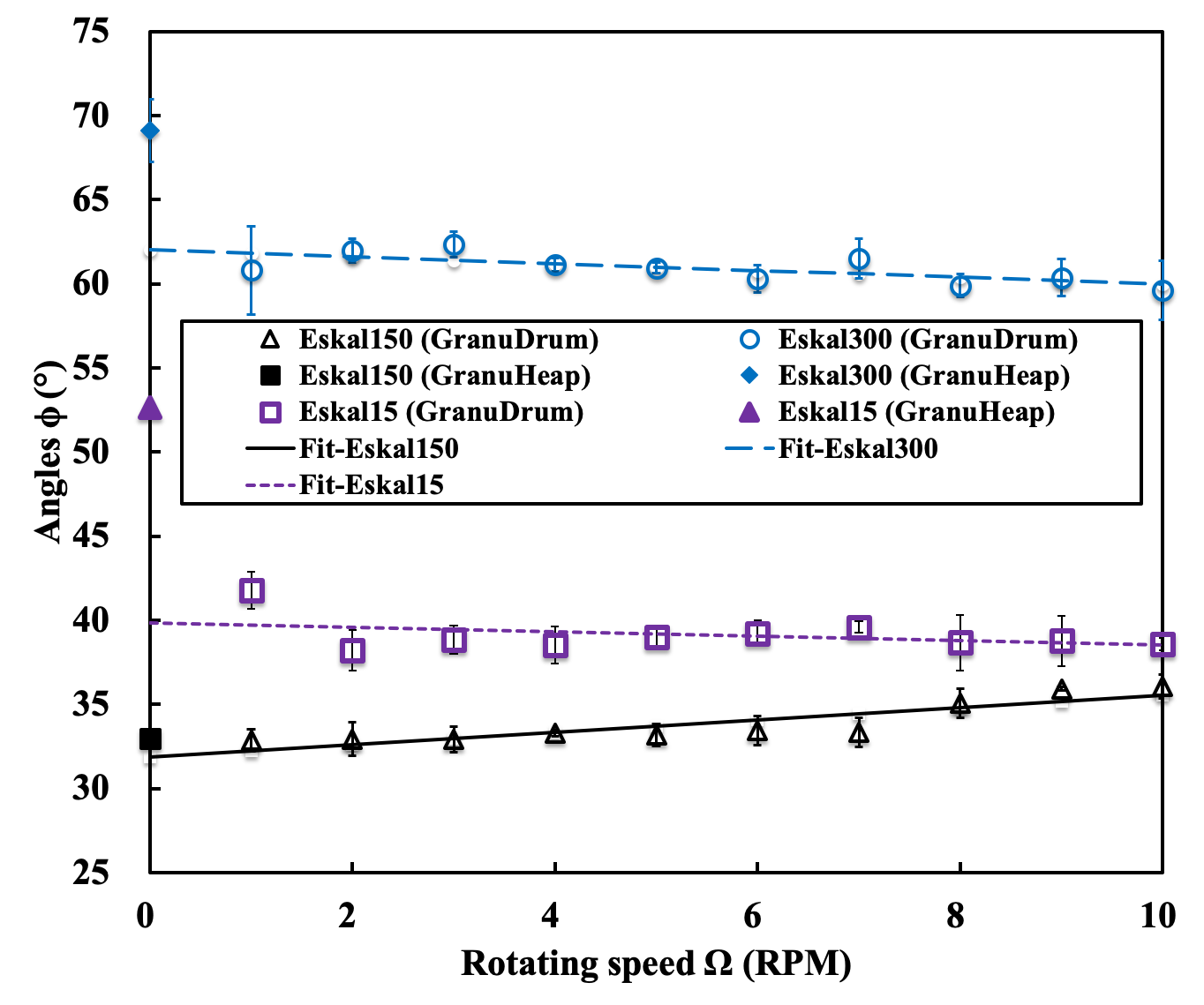}\label{Eskal150-300DrumHeapAngle}
\caption{Angle of repose measured with the GranuHeap (plain symbols) and the flow angles for different rotating speeds measured with the GranuDrum (open symbols). A linear regression allows to extrapolate the angle at 0 rpm from GranuDrum data with $\phi = \phi_{\Omega0} + \phi_{\Omega1} \Omega$, with $\phi_{\Omega0} = 32^\circ, 40^\circ, 62^\circ$ and $\phi_{\Omega1} = 0.37, -0.13, -0.20$ for Eskal150, Eskal15 and Eskal300, respectively.}
\label{fig-sec5:Eskal150-300DrumHeapAngle}
\end{figure}

The flow angles of our three limestone powders at different rotating speeds are measured with the GranuDrum and plotted in Fig.\ \ref{fig-sec5:Eskal150-300DrumHeapAngle}. As a function of the rotating speed, the flowing angle increases for the free flowing Eskal150 and decreases for the cohesive Eskal300. This behavior is also commonly seen for other powders \cite{lumay2012measuring}. The increase with rotating speed for non-cohesive granular material is due to the inertial effect, while the decrease for cohesive powder is due to stronger aeration at higher rotating speeds. A linear regression allows us to extrapolate to the angle at 0 rpm and we obtain $\phi_{dyn}=32^\circ$ for non-cohesive Eskal150, $\phi_{dyn}=43^\circ$ for slightly cohesive Eskal15 and $\phi_{dyn}=62^\circ$ for cohesive Eskal300. In addition, we also plotted in Figure \ref{fig-sec5:Eskal150-300DrumHeapAngle} the three angles of repose measured with the GranuHeap at zero rotating speed for comparison. For the free flowing Eskal150, the angle of repose measured from GranuHeap is comparable to the extrapolated flow angle at 0 rpm. However, for the cohesive powders Eskal15 and Eskal300, the angles of repose measured from the heaps,  $\phi_{sta}$, are considerably higher than the angle extrapolated from the GranuDrum data. This difference can be explained by the existence of two angles measured respectively before and after the slope instability (avalanches), which are named upper and lower angle \cite{cheng2017difference}. The angle of repose measured in Sec.\ \ref{subsec:results-heap} represents the highest stable angles that Eskal300 and Eskal15 could ever reach (upper angle) while the flow angles stay always between the upper and lower angles. Some previous studies \cite{lajeunesse2004spreading,liu2011measuring,al2016correlation} revealed that several influencing factors of using the hollow cylinder preparation method with establishing a different history for the powder: stratification, interface friction angle (which is the friction angle between the base and the granular material), lifting velocity, cylinder size, base roughness, granular material mass and height of the material in the cylinder. As the lifting velocity, material mass and material height increase, the angle of repose decreases. However, when the roughness of the base increases, the angle of repose also increases. This could possibly explain the higher values we measured here as our lifting velocity (5 mm/s) and material mass (height control with low bulk density) are both low. If we increase the lifting velocity or the initial filling mass, the measured angle of repose will become lower. However,  our main goal here is to reach the static free surface limit without varying the standard testing protocol, therefore we keep the measurement conditions as it is.

Note that different PSDs could lead to the changes of powder flowability. For cohesive Eskal300, the size range is between 1 and 10 \textmu m, thus we do not expect such a low energy input (zero to very low confining stress) will lead to any attrition effect. Instead, the agglomeration due to centrifugal force at high rotation speed could lead to significant change in the angles. For moderate cohesive Eskal15, the median particle size is almost 10 times larger than that of Eskal300, therefore the expected attrition/agglomeration effect is also negligible. Moreover, here we focus on the steady state friction which is the angle that does not vary with time or further deformation. In this study, we try to avoid going to too high rotation speeds as our focus is the quasi-static limit state. For free flowing Eskal150, the steady state angle of internal friction stays almost constant in all the tests performed, thus no change of PSD is expected.

\FloatBarrier

\subsection{From Small to Large Confining Stress}
\label{subsec:results-join}

After confirming the repeatability of each test, we come back to the main focus of this study: linking different flow regimes, not only from dynamic to static, but also from moderate to low and almost no confining stress. Our first step is to explore the lowest confining stresses of the Schulze ring shear tester in the quasi-static flow regime and extrapolate the steady state angle of internal friction to zero confining stress, which is relevant to a free surface flow. The second step is to evaluate the values of effective confining stress for both (static) GranuHeap and (dynamic) GranuDrum. Results can then be presented in a unique comprehensive plot showing the dependence of the steady state angle of internal friction $\phi$ on the confining stress $\sigma$ for the three tests. 

\begin{figure}[!ht]
 \centering
\includegraphics[scale=0.4,angle=0]{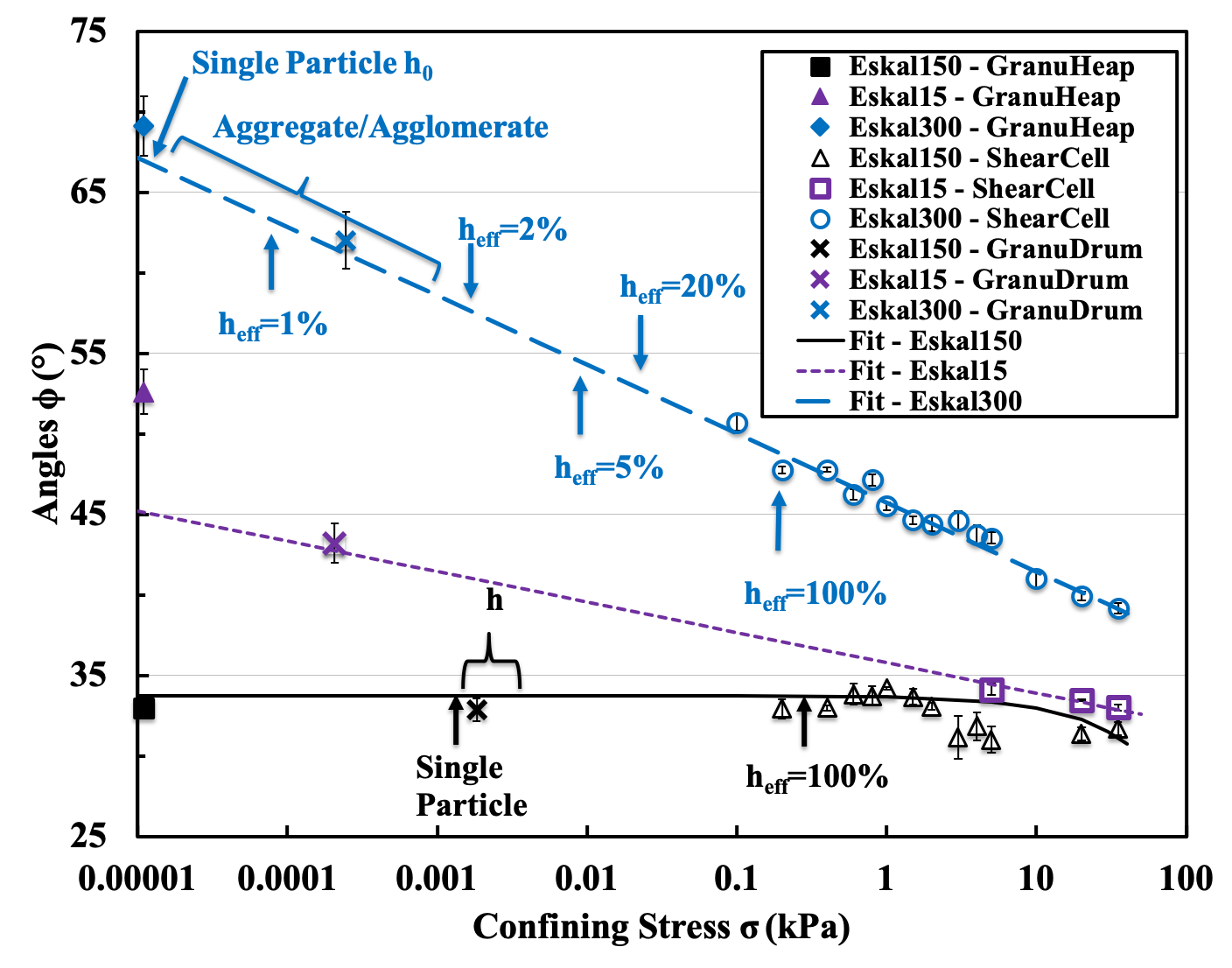}\label{chapter4-drum-RST-fitting}
\caption{Angles, $\phi$, from different types of tests as a function of confining stress, $\sigma$ for Eskal150 (138 \textmu m) and Eskal300 (2.2 \textmu m) in semi-log scale. The confining stresses refer to the normal stress at pre-shear in the ring shear test, and the estimated stresses from the weight of a single particle layer and effective flowing depth of powder in GranuHeap and GranuDrum, respectively.
Lines are the fitting to the shear test data: black linear regression line, $\phi = \phi_0 (1 - \sigma/\sigma_{\phi}$), with the limit ($\sigma \to 0$) angle $\phi_0 = 33.73^\circ$ and characteristic stress $\sigma_{\phi} = 452$ kPa for Eskal150; dashed purple line, $\phi = \phi_1 - \Delta\phi \mathrm{log}(\sigma/\sigma_{1})$, with $\phi_1 = 35.78^\circ$, $\Delta\phi =  0.82^\circ$ and $\sigma_{1} = 1$ kPa for Eskal15;
dashed blue line, $\phi = \phi_1 - \Delta\phi \mathrm{log}(\sigma/\sigma_{1})$, with $\phi_1 = 45.85^\circ$, $\Delta\phi = 1.86^\circ$ and $\sigma_{1} = 1$ kPa for Eskal300. Arrows indicate the estimated effective confining stresses assuming a single particle layer (changes a lot, since $\propto d_p^3$) or effective flowing depth  $h_{eff} = h/r$ in the rotating drum (varies a little, since $\propto \rho_{bulk}$). Aggregate/agglomerate refers to the clumps formed/destroyed due to the cohesiveness among powder particles.}
\label{fig-sec5:drum-RST-fitting}
\end{figure}

In Fig.\ \ref{fig-sec5:drum-RST-fitting}, the steady state angle of internal friction measured by the Schulze ring shear tester are plotted against the confining stress for the selected Eskal powders. The confining stress axis is shown in logarithmic scale in order to represent better the low stress range. We have fitted Eskal150 data using a linear regression and the Eskal300 data using a logarithmic decay; those fitted functions have been extended towards the very low stress regime. In the same figure, we have also indicated the values of angle of repose and the extrapolation of the flow angle from the rotating drum at 0 rpm for both Eskal150 and Eskal300. Note that here we have also included the data at larger pre-shear stresses ($\sigma > 2$ kPa) from the previous study \cite{shi2018effect} for the sake of completeness and validation.

For the free flowing Eskal150, the linear regression in the low confining stress regime ($\sigma \leq 2$ kPa) stays almost constant. This behaviour is mainly dominated by the surface properties of the primary particles, e.g., shape, roughness, and thus almost not influenced by the confining stress. On the other hand, if the confining stress becomes larger ($\sigma > 2$ kPa), the fitted line decreases slightly with the confining stress. The higher confining stress is a possibility to remove the factors dominant at low stresses, e.g., particles are rearranged to reduce the porosity or particles are more compressed towards each other to form contact flattening, and thus reduce the effect from particle surface irregularities. When we compare the Schulze ring shear tester data to the other two testers for Eskal150, both angle of repose $\phi_{sta}$ (black) and flow angle $\phi_{dyn}$ at 0 rpm (black Drum-0 rpm) match well with the prediction (black fitting line) from the Schulze ring shear tester data. This material is free flowing and insensitive to the confining stress in the low pressure range of interest. 

For cohesive Eskal300, the best fit is obtained by a reducing logarithmic decay. 
This decreasing trend with confining stress is expected as powders normally flow better in the larger confining stress regime.  The reason is that larger confining stress leads to larger rearrangements, plastic deformations and possibly contact flattening. This reduces the influences from surface roughness and geometrical interlocking, and thus results in a reduction of flow resistance. 
For GranuHeap tests, we have used the principle of single particle layer $h_0$ in the method explained in Sec.\ \ref{sec:experimental} to estimate the effective confining stresses. Furthermore, we have also tested this reducing logarithmic decay using another slightly cohesive Eskal15 powder and the ring shear tests data are taken from our previous study \cite{shi2018effect}. Although the shear test measurement points are limited compared to the cohesive Eskal300, the proposed decay looks plausible, but more data are needed in the future. The fitted function describes very well the results from both GranuDrum and ShearCell tests and further supports the observation of cohesive Eskal300, but the angle of repose stays higher and deviates from the fitted line. 
 
To estimate the effective confining stresses in GranuDrum, the second principle of effective flowing depth $h$ (see Sec.\ \ref{sec:experimental} for details) is used to evaluated at different depths $h$ inside the powder bed considering the hydro-static pressure $\sigma = \rho_{bulk} g h$, where $\rho_{bulk}$ is the powder bulk density and $g$ is gravitational acceleration. These points given by the estimated confining stress are shown with arrows in Fig.\ \ref{fig-sec5:drum-RST-fitting}. We do not go beyond 20\% of the drum radius, as the flow is sometimes more like a snow ball rolling down an inclined plane and the flow angle measurement becomes questionable. As there is no any agglomeration expected for free flowing Eskal150, we estimate the confining stress ($\sigma \approx 0.00183$ kPa) using the median particle size as the effective flowing depth ($h$ = 138 \textmu m). For Eskal15, a similar estimation ($h$ = 19 \textmu m) is used to estimate the confining stress ($\sigma \approx 0.00021$ kPa). Whereas for strongly cohesive Eskal300, the average estimated agglomerate size ($h \approx 50$ \textmu m) is used to extract the effective confining stress ($\sigma \approx 0.00025$ kPa).

 The angle of repose of Eskal300 (blue) is very close to the prediction of the confining stress from single particle layer and also agrees well with the steady state angle of internal friction of Eskal300 as extrapolated by the fitting of shear test data. The flow angle of GranuDrum at 0 rpm (blue solid diamond) is also plausible and sits well on the dashed line. Finally, the confining stress estimation at the bottom of the drum ($h_{eff} = 100 \%$) for both Eskal150 and 300 are also given on the same figure. They reach into the data points obtained by shear cell measurements at low confining stress levels. The good agreement of the three types of tests shows the possibility of extending the instrumental measuring limits by means of an accurate comparison of different types of tests.

\section{Conclusion and Outlook}
\label{sec:conclusionandoutlook}

In this study, we have examined the flow behaviour of three non-cohesive (Eskal150) and cohesive (Eskal15, 300) limestone powders in three  characterization tests: GranuHeap (angle of repose), Schulze ring shear test (steady state angle of internal friction) and GranuDrum (flow angle). Tests at various low levels confining stresses are performed in the Schulze ring shear tester and the results are extrapolated towards almost zero confining stress with empirical laws. This offers the possibility to extend the low limit of the confining stress and thus link to the other types of tests, specifically dynamic tests. To our knowledge, there is no similar study done before, although those tests have been used in the powder technology community for a very long time.

The angle of repose of free flowing Eskal150 ($d_{50} = 138$ \textmu m) measured with GranuHeap is much lower than the angle of repose of Eskal300 ($d_{50} = 2.2$ \textmu m), which indicates that higher cohesion also correlate with higher angles (shear resistance) and higher non-linearity for the same material, but smaller size.

The flow angle of Eskal150 measured with GranuDrum increases at large rotation speeds, while a systematic decrease is observed for both Eskal15 and Eskal300. This is expected since common flow behaviour of free flowing and cohesive powders are dominated by inertial effects and aeration, respectively. The extrapolation of the flow angle of Eskal150 to 0 rpm agrees well with the measured angle of repose. However, the extrapolations of the flow angle of Eskal15 and Eskal300 to 0 rpm are significantly lower than the angle of repose. Indeed, the angle of repose is the highest stable angle while the flow angle is between the angle of repose and the angle after an avalanche. 

The steady state angle of internal friction obtained by quasi-static ring shear tests is found to be a function of confining stress. The data of free flowing Eskal150 are fitted well by a linear regression whereas the cohesive Eskal300 is well described by a logarithmic decay. These two empirical laws allow us to predict very low confining stresses that are comparable with the other two types of tests. For free flowing Eskal150, all three tests agree very well with each other. For the cohesive Eskal300, the estimation of the effective confining stress becomes very difficult in the free surface tests. Nevertheless, the results of the slightly cohesive Eskal15 from three different tests are described also very well with the reducing logarithmic decay function, confirming the validity of the approach we used here. Our method opens new perspectives in the field of powder characterization and for measurement interpretation. Cohesive powders in industrial process lines with small to moderate stresses (1-100 Pa) might suffer from unusually large bulk friction. The empirical logarithmic stress dependence of the steady state angle of internal friction allows to close the gaps, where measurements are difficult.

In future, the applicability of the proposed empirical laws should be further checked by including new materials or alternative testing techniques. Also a more detailed study of the effective flowing layer depth in a rotating drum for cohesive powders is needed as well as an inside view into both heap and shear cell to understand the differences.

\section*{Acknowledgement}
We would like to thank for the financial support through the ``T-MAPPP" project of the European-Union-Funded Marie Curie Initial Training Network FP7 (ITN607453), in which the Schulze ring shear measurements were carried out. GranuDrum and GranuHeap measurements were conducted in the framework of the ``PowderReg'' project, funded by the European programme Interreg VA GR within the priority axis 4 ``Strengthen the competitiveness and the attractiveness of the Grande R\'egion / Gro{\ss}region''. The help from R. Cabiscol on making the SEM images and shear test experiments is also greatly acknowledged.

\section*{Conflict of interest}
The authors declare that there is no conflict of interest.

\section*{References}

\bibliography{library.bib}

\clearpage

\begin{table*}[ht]
\caption{Material parameters of the limestone samples. The initial bulk density represents bulk density from raw materials, as provided by the manufacturer.}\label{tablematerialEskal}
\centering
\scalebox{1.0}{
 \begin{tabular}{*{12}{c}}
    \hline
     \bf{Property} &  & \bf{Unit} & \bf{Eskal 300} &\bf{Eskal15} &\bf{Eskal150} \\ [0.4ex]
    \hline 
      & $d_{10}$& \textmu m & 0.78 & 12 &  97  \\ [0.4ex]
    Particle Size&  $d_{50}$ & \textmu m & \bf{2.22}  & \bf{19} &  \bf{138} \\ [0.4ex]
    & $d_{90}$ & \textmu m  & 4.15 & 28  & 194  \\ [0.4ex]
    Span & ($d_{90}$-$d_{10}$)/$d_{50}$ & [-] & 1.52 &  0.84 & 0.70  \\ [0.4ex]
     Particle density & $\rho_\mathrm{p}$ & kg/m\textsuperscript{3} & 2853 & 2737 & 2761\\ [0.4ex]
    Moisture content  & $w$ & $\%$ &  0.9  & 0.9&   0.9 \\ [0.4ex]
   Roundness & $\Psi$ & [--] & 0.75 &  0.48 & 0.88 \\ [0.4ex]
   Initial bulk density & $\rho_0$ & kg/m\textsuperscript{3} & 540  & 1110 &  1370 \\ [0.4ex]
    \hline
  \end{tabular}
  }
\end{table*}

\end{document}